\documentstyle[aps,preprint,prl]{revtex}

\begin{document}

\draft
\preprint{\begin{tabular}{r}FTUAM 96/15\\UFIFT-HEP-96-8\end{tabular}}
\title{A dark matter solution from the supersymmetric axion model}
\author{Sanghyeon Chang$^1$\cite{a} and Hang Bae Kim$^2$\cite{b}}
\address{$^1$Institute for Fundamental Theory,
             University of Florida, Gainesville, FL 32611, USA\\
             $^2$Departamento de F{\'\i}sica Te\'orica C-XI,
             Universidad Aut\'onoma de Madrid,\\
             Cantoblanco, 28049 Madrid, Spain}
\maketitle
\begin{abstract}
We study the effect of the late decaying saxino (the scalar superpartner of
the axion) and find out that there is a possible dark matter solution from a
class of supersymmetric extensions of the invisible axion model.
In this class of models, the saxino which decays into two axions acts as the
late decaying particle which reconciles the cold dark matter model
with high values of
the Hubble constant.  Recent observations of the Hubble constant are
converging to $H_0=70\!-\!80\,{\rm km}\,{\rm sec}^{-1}\,{\rm Mpc}^{-1}$, which
would be inconsistent with the standard mixed dark matter model.  This class
of models provides a plausible framework for the alternative cold dark matter
plus late decaying particle model, with the interesting possibility that both
cold dark matter and the extra radiation consist of axion.
\end{abstract}


Cosmological and astrophysical data accumulated in past several years made
it possible to test the theories of large scale structure formation in our
universe.  The COBE-DMR discovery of cosmic microwave background(CMB)
anisotropy provided the means for the accurate normalization of theories of
structure formation.  The study of galaxy correlations supplied complementary
data on smaller scales.  With the scale invariant initial spectrum, they
manifested that the pure hot dark matter model cannot explain the small scale
structure of the universe and the pure cold dark matter(CDM) model shows best
fit at the values $\Omega_0h=0.25$ \cite{PD94}
($\Omega_0=$(the present energy density)/(the critical energy
density) and $h=$(the present value of the Hubble constant)/$100\,{\rm
km}\,{\rm sec}^{-1}\,{\rm Mpc}^{-1}$), which is much smaller than the
presently favored values.  The mixed dark matter(MDM) model attracted broad
attention because it gives good fit for $\Omega_\nu=0.2$ and $\Omega_{\rm
CDM}=0.8$ at $h=0.5$ \cite{MDM}.

For better test of models, however, the accurate values of $\Omega_0$ and $h$
have been required.  The observed value of $\Omega_0$ is ranging from 0.2 to
$\sim1$, still plagued by large systematic uncertainty.  Several recent
investigations of the Hubble constant tend to give higher values, in the range
of $h=0.7\!-\!0.8$ \cite{Hubble_constant}.  If we stick to the theoretical
prejudice, $\Omega_0=1$, we have at least two serious problems. First, we
cannot fit the power spectrum curve even in the MDM model.  Second, it would
be inconsistent with the estimated lower bound on the age of the universe from
the oldest globular clusters.  An $\Omega_0=1$ universe has an age of only
$t_0=6.5h^{-1}\,{\rm Gyr}$, giving $t_0<9.3\,{\rm Gyr}$ for $h>0.7$.  On the
other hand, the observed value of the age of the oldest globular cluster is
around 15 Gyr\cite{GC}.  Still there are a few possible way to relax this
bound, but it seems not to be less than 11 Gyr\cite{GC2}.  Assuming that the
data from the globular cluster can be relaxed by some mechanism and
considering error bars in Hubble constant observations, $h=0.6$ would be
marginally allowed.  However, the standard MDM model is inconsistent with
large scale structure data even for $h=0.6$.

Introduction of the small cosmological constant corresponding to
$\Omega_\Lambda=0.7\!-\!0.8$ alleviates the universe age crisis mildly and can
revive good features of the CDM model with $\Omega_\Lambda+\Omega_{\rm CDM}=1$
(the CDM$+$$\Lambda$ model).  However, such a small value of cosmological
constant is still a theoretically knotty subject.  One way to keep the
cosmological constant to be zero and maintain good features of the CDM model
is introducing the late decaying particle(LDP) (the CDM$+$LDP model)
\cite{KK95}.  In the CDM$+$LDP model, the LDP decays into very light and
weakly interacting particles in the late stage of universe.  This new
radiation dominates the radiation energy of universe and would delay the
beginning of the matter dominated epoch, which is necessary for the CDM model
to be consistent with structure formation data even at the high values of $h$.
But there are severe restrictions on the mass, lifetime and interactions of
the LDP coming from structure formation data and preserving the successful
predictions of nucleosynthesis and the CMB spectrum.  We find that these can
be met for the axion supermultiplet in supersymmetric extensions of the
well-known axion model.  It can provide cold dark matter with the late
decaying particle consistently in a class of models.

The axion is originally introduced to solve the strong CP problem.  
The axion model has an interesting
cosmological effect.  The axion produced from the initial vacuum misalignment
has very small momentum  and therefore is a good candidate of the cold dark
matter.

In the supersymmetric extension of the axion model, the axion
supermultiplet($\Phi$) consists of the axion($a$), its real scalar
superpartner saxino($s$), and the fermionic superpartner axino($\tilde a$).
The cosmological impacts of these two additional weakly interacting particles  
were studied before \cite{RTW91,Kim91}.  In this paper, we present unappreciated
possibility of saxino cosmology, in which the axino decays to two axions and
acts as a LDP.

After supersymmetry breaking, the axino and the saxino get their masses.
In global supersymmetry, they remain massless at the tree-level.
In the supergravity model with supersymmetry broken in the hidden sector, 
they are
expected to gain masses of order $m_{3/2}$.  However the axino and saxino
masses are dependent on the specific forms of the axion sector superpotential
as well as the K\"ahler function in the context of supergravity
\cite{GY92}.  A model-dependent analysis is necessary to evaluate to the
masses of axion supermultiplet.  With the minimal K\"ahler function, the
saxino mass is roughly of order of $m_{3/2}$ in most cases.  In no scale
supergravity, it remains massless at the tree level and gains radiative
corrections of order $10\sim100{\rm MeV}$.

The effective Lagrangian describing the interactions of the axion
supermultiplet is given by \cite{CL95,Kim91}
\begin{eqnarray}
{\cal L} &=&
\left.\sum_iv_i^2\exp\left[q_i(\Phi+\overline{\Phi})/F_a\right]\right|_D +
\left.\frac{\alpha_c}{16\pi F_a}\Phi W_\alpha W^\alpha\right|_F
\nonumber\\ &=&
\left(1+\frac{\sqrt2x}{F_a}s\right)\left(
\frac12\partial_\mu a\partial^\mu a +
\frac12\partial_\mu s\partial^\mu s +
i\overline{\tilde{a}}\gamma_\mu\partial^\mu\tilde{a} \right) 
\nonumber\\ &&
+\frac{\alpha_c}{8\pi F_a}\left(
aF^a_{\mu\nu}\tilde{F}^{a\mu\nu} 
+\tilde{a}\gamma_5\sigma_{\mu\nu}\lambda^aF^{a\mu\nu} 
+sF^a_{\mu\nu}F^{a\mu\nu} + \cdots \right)
-\frac{x}{F_a}\partial_\mu a\;\overline{\tilde{a}}\gamma^\mu\tilde{a} +
\cdots .
\end{eqnarray}
where $F_a$ is the axion decay constant and $\alpha_c$ is color coupling
constant.  The model-dependent parameter $x$ is given by
$x=\sum_iq_i^3v_i^2/F_a^2$ where $q_i$ and $v_i$ are the $U(1)_{\rm PQ}$
charges and the VEVs of the fields in the axion sector.  It is of order 1,
in general.  At present particle phenomenology, astrophysical and cosmological
observations restrict the range of the axion decay constant to be 
\begin{equation}
10^{10}{\rm GeV} \lesssim F_a \lesssim 10^{12}{\rm GeV}.
\label{axion_decay_constant}
\end{equation}

{}From these couplings we obtain the saxino decay widths to two different
channels
\begin{eqnarray}
\Gamma_{s\rightarrow2g} &=& \frac{\alpha_c^2m_s^3}{128\pi^3F_a^2}, \\
\Gamma_{s\rightarrow2a} &=& \frac{x^2m_s^3}{64\pi F_a^2},
\end{eqnarray}
where $m_s$ is the saxino mass.
Depending on the model dependent parameter $x$, the main decay mode can
be changed
and the effect of the saxino decay appears quite different.
When $s\rightarrow2g$ is the main decay mode, the lifetime is given by
\begin{equation}
\tau_s = 2.6\times10^6{\rm sec}\ 
\left(\frac{\alpha_c}{0.1}\right)^{-2}
\left(\frac{F_a}{10^{11}{\rm GeV}}\right)^2
\left(\frac{100{\rm MeV}}{m_s}\right)^3.
\end{equation}
The resulting gluons thermalize and dump some amount of entropy to thermal
bath \cite{Kim91}.  Previous studies on the saxino decay have been
focused on the saxino heavier than $10\,{\rm GeV}$ because it has been
believed that the saxino should decay before nucleosynthesis. The reason is
that the high energy radiation from the late decaying saxino would destroy the
light nuclei which were made during the nucleosynthesis era.  However when
$x\gg10^{-3}$, as expected in many models, the main decay mode of
the saxino is $s\rightarrow2a$ with the lifetime
\begin{equation}
\tau_s = 1.3\times10^3{\rm sec}\ x^{-2}\left(\frac{F_a}{10^{11}{\rm GeV}}
\right)^2\left(\frac{100{\rm MeV}}{m_s}\right)^3.
\label{saxino_lifetime}
\end{equation}
The crucial difference is that the resulting axions interact very weakly and
cannot be thermalized.  They do not dump entropy to thermal bath of the
universe and do no harm to the light nuclei made during nucleosynthesis.
However, their effect appears gravitationally through their energy density.
It affects the result of nucleosynthesis, which makes us consider the
different range of the saxino mass.

For further discussion, we need to know the relic abundance of the saxino.
There are two distinct contributions: thermal and non-thermal.
First we consider the thermal relic.  Without inflation,
the saxino would be hot thermal relic which decouples at the temperature,
as estimated by Rajagopal et al.\ \cite{RTW91},
\begin{equation}
T_{\rm dec} \simeq 10^{9}{\rm GeV}
\left(\frac{F_a}{10^{11}{\rm GeV}}\right)^2
\left(\frac{\alpha_c}{0.1}\right)^{-3}.
\label{dec}
\end{equation}
Then its relic density is given by
\begin{equation}
Y_s \equiv \frac{n_s}{s}
= 0.278\frac{g_{\rm eff}}{g_{*s(T_{\rm dec})}} \simeq 1.2\times10^{-3},
\label{saxino_abundance1}
\end{equation}
where $n_s$ is the number density of the saxino and $s=(2\pi^2/45)g_{*s}T^3$
is the entropy density.

In the inflationary scenario, the reheating temperature is an important
parameter in our discussion because the decoupling temperature of the saxino
is rather high.  At present, the relevant upper bound on the reheating
temperature comes from the gravitino production and is quoted as \cite{RS95}
\begin{equation}
T_{\rm reh}\lesssim10^9{\rm GeV}
\end{equation}
for reasonable values of the gravitino mass.  If we use the MSSM value
$\alpha_c(10^{11}{\rm GeV})\simeq1/20$ in the Eq.~(\ref{dec}), the reheating
temperature seems lower than the decoupling temperature.
However, in the heavy quark axion model, the heavy quark and saxino coupling
is about (mass of heavy quark)$/F_a$, which is much stronger than any other
interaction with saxino.  In this case, the axion will not decouple until the
heavy quark decouples from the heat bath.  Therefore the decoupling
temperature could be lower than reheating temperature.  Though the decoupling
temperature is greater than the reheating temperature, saxino could be
produced from the thermal bath through the reactions like
$q\bar{q}\leftrightarrow sg$ and $gg\leftrightarrow sg$.  Then the relic
density is estimated to be
\begin{equation}
Y_s \simeq 0.75\times 10^{-4}
\left(\frac{\eta\alpha_c^3}{10^{-4}}\right)
\left(\frac{F_a}{10^{11}{\rm GeV}}\right)^{-2}
\left(\frac{T_{\rm reh}}{10^{9}{\rm GeV}}\right),
\label{saxino_abundance2}
\end{equation}
where $\eta$ is phase space factor times the number of channels, and
correction from the thermal effects which is roughly of order 1.

As non-thermal relic, there can be a coherent oscillation of the saxino field.
The coherent oscillation begins when the supersymmetry breaking saxino mass
becomes comparable to the Hubble parameter.
We estimate the relic density from coherent oscillation to be
\begin{equation}
Y_s \simeq 5\times 10^{-8}
\left(\frac{m_s}{10{\rm MeV}}\right)^{-1/2}
\left(\frac{s_1}{10^{11}{\rm GeV}}\right)^2,
\label{saxino_abundance3}
\end{equation}
where $s_1$ is the initial value of the saxino when the oscillation begins.
The misalignment production is negligible for the saxino when $F_a <10^{13}$
GeV.

{}From the relic abundance of the saxino, we obtain strong bound on the saxino
mass comes from nucleosynthesis.  To preserve successful nucleosynthesis, the
additional energy density at the time of nucleosynthesis should be smaller
than, say, $\Delta N_\nu$ times the energy density of one neutrino species.
For the saxinos and decay produced axions, this corresponds to
$mY<1.3\Delta N_\nu(T_D/g_{*s(T_D)})$ for $T_D>T_{\rm NS}$ and $mY<1.3\Delta
N_\nu(T_{\rm NS}/g_{*s(T_{\rm NS})})$ for $T_D<T_{\rm NS}$ where
$T_D=0.55g_{*T_D}\sqrt{M_P\Gamma}$ is the temperature at decay time and
$T_{NS}\simeq0.8$ MeV the temperature at the time of nucleosynthesis.
Combining these, we obtain the axino mass bound
\begin{equation}
m_s > \frac{4{\rm TeV}}{x^2\Delta N_\nu^2}
\left(\frac{g_{*sD}}{100}\right)^{5/2}
\left(\frac{Y_s}{10^{-3}}\right)^2
\left(\frac{F_a}{10^{11}{\rm GeV}}\right)^2
\end{equation}
or
\begin{equation}
m_s < 107{\rm MeV}\,\Delta N_\nu\left(\frac{Y_s}{10^{-3}}\right)^{-1}.
\label{saxino_mass_upper_bound}
\end{equation}

If we adopt $\Delta N_\nu=0.3$ \cite{WSSOK91}, the upper bound in the
Eq.~(\ref{saxino_mass_upper_bound}) would be $30\,{\rm MeV}$ in the hot
thermal relic case.  Though the low reheating temperature weakens it, it is
still difficult to get such a small mass in models where the saxino gets its
mass at tree-level.  However in models like no scale supergravity models,
where the axino and the saxino have zero tree level masses and get masses by
radiative corrections, the $10\sim100\,{\rm MeV}$ mass of the saxino naturally
occurs in connection with the $\sim{\rm keV}$ order axino mass.

Then the mass, lifetime, decay products and relic abundance of the saxino
nicely fit with the requirements of the LDP.
The condition the late decaying particle should satisfy is \cite{KK95}
\begin{equation}
\left(\frac{\tau}{{\rm sec}}\right)\left(\frac{mY}{{\rm MeV}}\right)^2 \simeq 
    0.55\left[\left(\frac{h}{0.25}\right)^2 -1 \right]^{3/2}.
\end{equation}
For the hot saxino thermal relic, using the Eq.~(\ref{saxino_abundance1}),
it becomes
\begin{equation}
\frac{190}{x^2}\left(\frac{m_s}{10{\rm MeV}}\right)^{-1}
\left(\frac{F_a}{10^{11}{\rm GeV}}\right)^{2} \simeq 5.7
\end{equation}
for the $h=0.6$ case.  Considering the bounds on $F_a$ and $m_s$ (the
Eqs.~(\ref{axion_decay_constant}) and (\ref{saxino_mass_upper_bound})), this
can be met in models with $x\simeq1$ and $F_a\sim10^{10}\,{\rm GeV}$.
If the reheating temperature is lower than decoupling temperature of saxino,
we should use the Eq.~(\ref{saxino_abundance2}).  Then we obtain
\begin{equation}
\frac{7.0}{x^2}
\left(\frac{m_s}{100{\rm MeV}}\right)^{-1}
\left(\frac{F_a}{10^{11}{\rm GeV}}\right)^{-2}
\left(\frac{T_{\rm reh}}{10^{9}{\rm GeV}}\right)^{2}  
\simeq 5.7.
\end{equation}
In this case, depending on the reheating temperature $T_{\rm reh}$, the
condition can be satisfied in models with small $x$ and
$F_a\simeq10^{12}\,{\rm GeV}$.

The CDM$+$LDP model is distinguished from the CDM$+$$\Lambda$ model by the
existence of the intermediate matter domination era before the final
matter domination era. This temporary matter domination
occurs because the saxino dominates
the universe before it decays into the axions.
Due to the temporary matter domination, there is one more length scale 
\cite{KK95}
\begin{equation}
\lambda_{\rm EQ1} \simeq 8\times10^{-2}
\left(\frac{\rm MeV}{mY}\right)\,{\rm kpc}.
\end{equation}
Objects on this and smaller scales would collapse at high red shift and the
CDM$+$LDP model has more structure on these scales.  For hot saxino thermal
relic, this scale is
\begin{equation}
\lambda_{\rm EQ1} \simeq 6.7\,{\rm kpc}\,
\left(\frac{m_s}{10{\rm MeV}}\right)^{-1},
\end{equation}
and for regenerated saxinos,
\begin{equation}
\lambda_{\rm EQ1} \simeq 11\,{\rm kpc}\,
\left(\frac{m_s}{100{\rm MeV}}\right)^{-1}
\left(\frac{F_a}{10^{11}{\rm GeV}}\right)^{2}
\left(\frac{T_{\rm reh}}{10^{9}{\rm GeV}}\right)^{-1}.
\end{equation}
The existence of this scale might be better in explaining the abundance of the
high red shift quasars and globular clusters.

We should check one more constraint on the LDP.  Though the main decay mode of
the saxino is $s\rightarrow2a$, the radiative branching ratio should be
sufficiently suppressed not to spoil the nucleosynthesis result by
photodestruction and photoproduction of light nuclei. For $m_s>10\!-\!50\,{\rm
MeV}$ and $\tau_s>10^6\,{\rm sec}$, it is roughly $rm_sY_s<10^{-9}\,{\rm MeV}$
where $r$ is the radiative branching ratio \cite{EGLNS92}.  If the saxino is
lighter than $100$ MeV, the saxino cannot decay into hadrons, but can decay to
two photons or two leptons.  These decay channel is highly suppressed if the
saxino is much heavier than 1 MeV.  Decay to two photons is also suppressed by
$\alpha^2$.  Two lepton mode is suppressed by loop factor in the heavy quark
axion model.

Finally we discuss the possible CDM candidates in the above context.  Actually
any CDM is good.  But because the supersymmetric axion model has two
candidates, the axion and the axino, in itself, we discuss only about them.
Though the mass of the axion is very small ($m_a\simeq\Lambda_{\rm
QCD}^2/F_a$), the large amount of axions can exist in our universe in the form
of the coherent oscillation arising from the initial misalignment of the
vacuum \cite{caxion}.  Its present energy density grows as $F_a$ gets larger
and becomes as large as the critical energy density around $F_a=10^{12}\,{\rm
GeV}$.  The upper bound on $F_a$ comes from the critical energy density.  When
the reheating temperature is larger than the decoupling temperature of the
saxino, $F_a\sim10^{10}\,{\rm GeV}$ is necessary for the saxino to be the LDP.
But when the reheating temperature is lower, $F_a\simeq10^{12}\,{\rm GeV}$ is
desirable because the saxino can be the LDP even in models with small $x$.
Therefore, we have an interesting possibility that CDM consists of the axions,
and the saxino acts as the LDP, which decays to provide additional radiation
which consists again of axions.

The decoupling temperature and the relic number density of the axino are quite
similar to those of the saxino.  When $T_{\rm reh}>T_{\rm dec}$, the stable
$\sim2\,{\rm keV}$ mass axino can be so called warm dark matter
\cite{RTW91,GY92}.  When $T_{\rm reh}$ is lower, the corresponding mass is
\begin{equation}
m_{\tilde a}\simeq30\,{\rm keV}\,
\left(\frac{F_a}{10^{11}{\rm GeV}}\right)^2
\left(\frac{T_{\rm reh}}{10^{9}{\rm GeV}}\right)^{-1},
\end{equation}
and the axino becomes CDM.  Interestingly enough, this axino mass range for
which the axino is CDM and the saxino mass range for which the saxino is a LDP
simultaneously occur in models where the axino and the saxino get their masses
by radiative corrections \cite{RTW91,GY92}.

In conclusion, we point out that the axion supermultiplet which solves the
strong CP problem can provide cold dark matter with $\Omega_{\rm CDM}=1$ and
the late decaying particle simultaneously within the presently allowed
parameter range if the $10\,{\rm MeV}\sim1\,{\rm GeV}$ saxino decays mainly into
two axions.  All the discussions are based on the flat, cosmological constant
free universe.  In this context, this model is one of a few possible models
which satisfy the large scale structure and the large Hubble constant.  It can
survive the universe age problem if we allow the marginal value 11 Gyrs.  If
the cosmological and astrophysical observations are refined, the validity of
this model will be proven in the near future.

\acknowledgments

We thank K. Choi, E.J. Chun and P. Sikivie for helpful discussions.
This work was supported in part by KOSEF and IFT at University of
Florida(S.C.) and the Ministerio de Educaci\'on y Ciencia under research
grant(H.B.K.).

\end{document}